\documentstyle[twocolumn,aps,epsf]{revtex}

\def\d{{\rm d}}

\def\vector#1{{\bf #1}}

\def\vk{{\vector k}}

\def\dps{\displaystyle}

\def\Tc{{T_{\rm c}}}

\def\TMTSFX{\mbox{${\rm (TMTSF)_2X}$}}
\def\TMTSFClO{\mbox{${\rm (TMTSF)_2ClO_4}$}}
\def\TMTSFPF{\mbox{${\rm (TMTSF)_2PF_6}$}}

\def\BEDTTTFX{\mbox{${\rm (BEDT}$-${\rm TTF)_2X}$}}
\def\BETSFeCl{\mbox{${\rm (BETS)_2FeCl_4}$}}

\def\RSGCO{\mbox{${\rm RuSr_2GdCu_2O_8}$}}

\def\hsp#1{\hspace{#1ex}}

\def\Tc{{T_{\rm c}}}

\def\dblesum#1#2{\sum_{\stackrel{\scriptstyle #1}{\scriptsize #2}}}

\def\HP{{H_{\rm P}}}

\begin{document}
\draft

\twocolumn[\hsize\textwidth\columnwidth\hsize\csname 
@twocolumnfalse\endcsname

\title{Enhancement of the Upper Critical Field and a Field-Induced \\
Superconductivity in Antiferromagnetic Conductors 
}

\author{Hiroshi Shimahara}



\address{
Department of Quantum Matter Science, ADSM, Hiroshima University, 
Higashi-Hiroshima 739-8530, Japan
}

\date{October 15, 2001}

\maketitle

\begin{abstract}
We propose a mechanism by which the paramagnetic pair-breaking effect is 
largely reduced in superconductors with coexisting antiferromagnetic 
long-range and short-range orders. 
The mechanism is an extension of the Jaccarino and Peter mechanism to 
antiferromagnetic conductors, but the resultant phase diagram is quite different. 
In order to illustrate the mechanism, we examine a model which consists of 
mobile electrons and antiferromagnetically correlated localized spins 
with Kondo coupling between them. 
It is found that for weak Kondo coupling, 
the superconductivity occurs over an extraordinarily wide region 
of the magnetic field including zero field. 
The critical field exceeds the Chandrasekhar and Clogston limit, 
but there is no lower limit in contrast to the Jaccarino and Peter mechanism. 
On the other hand, for strong Kondo coupling, 
both the low-field superconductivity 
and a field-induced superconductivity occur. 
Possibilities in hybrid ruthenate cuprate superconductors and 
some organic superconductors are discussed. 
\end{abstract}

\pacs{
}



]

\narrowtext

Recently, superconductivity at high fields exceeding 
the Chandrasekhar and Clogston limit~\cite{Cha62,Clo62} 
(Pauli paramagnetic limit $\HP$) has been examined 
by many authors~\cite{Shi94a,Shi98,Shi97b,Man00,Leb99,Yan00,Shi00a}, 
in connection with experimental data for organic 
superconductors~\cite{Lee97,Ish00,Sin00,Zuo00,Shimo01,Uji01}. 
In \TMTSFX \hsp{0.25} and $\kappa$-\BEDTTTFX \hsp{0.25} compounds 
in parallel magnetic fields, 
the upper critical field exhibits an upturn ($\d^2 H_{\rm c2}/\d T^2 > 0$) 
at low temperatures, exceeding the value of $\HP$ 
estimated from the zero-field transition temperature $\Tc^{(0)}$ 
by a simplified formula 
$H_{\rm P} \sim 1.86 \, \Tc^{(0)}$~\cite{Lee97,Ish00,Sin00,Zuo00,Shimo01}. 
On the other hand, in a $\lambda$-\BETSFeCl \hsp{0.25} compound, 
a field-induced superconductivity was observed at high fields, 
while at zero field it is an antiferromagnetic insulator~\cite{Uji01}.

There are several possible mechanisms 
to explain the critical fields higher than $\HP$. 
For example, triplet superconductivity, 
the Fulde-Ferrell-Larkin-Ovchinnikov state 
(the FFLO state or LOFF state)~\cite{Ful64,Lar64}, 
spin-orbit coupling, 
and a strong coupling effect have been examined. 
Among them, the upturn of the critical field can be explained 
by triplet superconductivity 
with dimensional crossover~\cite{Leb99}, 
and also by the FFLO state~\cite{Shi94a,Shi98,Shi97b}. 
The pairing symmetries in organic superconductors seem to be still 
controversial~\cite{Shi00c}.

In triplet superconductors, since the orbital pair-breaking effect 
is largely suppressed in a parallel magnetic field, 
the critical field of a parallel spin pairing is much higher than $\HP$. 
The upturn and a reentrant superconducting transition are predicted 
by a theory of dimensional crossover~\cite{Leb99}.

The microscopic origin of the triplet superconductivity has been 
studied by many authors. 
For example, 
on the analogy of the superfluid $^3$He, 
a paramagnon theory based on the ferromagnetic fluctuation 
appears to be a natural explanation, 
but many of the organics are in proximity to the antiferromagnetic phase. 
This discrepancy does not exist in phonon mechanisms of the triplet 
superconductivity. 
The anisotropic components of the pairing interactions are rather large 
due to the weak screening effect in the layered systems~\cite{Shi01}. 
Hence, reduction of the $s$-wave pairing interaction 
due to short-range Coulomb repulsion may give rise to 
triplet superconductivity~\cite{Shi01,Fou77}.

The pairing interactions mediated by antiferromagnetic fluctuations 
may also be a mechanism of triplet superconductivity. 
The pairing interactions mediated by antiferromagnetic fluctuations 
include attractive triplet channels~\cite{Shi00b}. 
Hence, suppression of the singlet superconductivity 
due to some additional mechanism, such as inter-site Coulomb repulsion 
or a magnetic field (if it is applied), 
may give rise to triplet superconductivity~\cite{Shi00b}.

On the other hand, in singlet superconductors, 
the FFLO state is a candidate for the mechanism 
of the high-field superconductivity 
and the upturn of the upper critical field. 
Recently, it was observed 
in $\kappa$-${\rm (BEDT}$-${\rm TTF)}_2{\rm Cu[N(CN)_2]Br}$ 
that the upturn has a tendency to disappear when the 
direction of the magnetic field is slightly tilted~\cite{Ohm01}. 
This behavior is consistent with a theoretical prediction based on 
the FFLO state. 
To obtain direct evidence of the FFLO state, observation of the spatial 
structure of the gap function by scanning tunneling spectroscopy (STS) 
would be useful~\cite{Shi98,Shi97b}. 
Purely quantitative arguments for the critical field are difficult 
because detailed information on the Fermi surface structure 
and the interactions is necessary 
for accurate estimation~\cite{Shi99,Shi00a,Mat94}. 
Subdominant interactions presumably exist in anisotropic 
superconductors~\cite{Shi01,Shi00b}.

Furthermore, 
it should be noted that in strong coupling superconductors, 
the observed critical fields are larger than the values of $\HP$ 
expected from $\Tc^{(0)}$ in weak coupling theory, 
because the ratio $\Tc^{(0)}/\Delta_0$ is reduced 
by a strong coupling effect, 
where $\Delta_0$ is the BCS gap at $T = 0$ and $H = 0$.

Jaccarino and Peter proposed a mechanism of ultrahigh-field 
superconductivity~\cite{Jac62}. 
They pointed out that 
the exchange field created by the polarized rare earth spin cancels 
the magnetic field in certain ferromagnetic metals. 
The Jaccarino and Peter mechanism explains the field-induced 
superconductivity in ${\rm Eu_xSn_{1-x}Mo_6S_8}$~\cite{Meu84} 
and probably that in the $\lambda$-\BETSFeCl \hsp{0.25} 
compound~\cite{Uji01}. 
An observed temperature dependence of the lower critical field of 
the high-field superconductivity in $\lambda$-\BETSFeCl~\cite{Uji01b} 
can be explained by the FFLO state in combination with the Jaccarino 
and Peter mechanism.

In this paper, we propose a mechanism by which 
the paramagnetic pair-breaking effect is largely reduced in 
superconductors with coexisting antiferromagnetic long-range and 
short-range orders. 
The present mechanism is an extension of the Jaccarino and Peter mechanism 
of ferromagnetic metals to antiferromagnetic metals. 
The origin of the enhancement of the critical field is due to 
a compensation effect similar to that of the Jaccarino and Peter 
mechanism. 
However, the resultant phase diagram is quite different 
for low fields or weak exchange fields. 
Possible candidates of the present mechanism may be found in some of 
the layered superconductors in proximity to the antiferromagnetic phase.

In order to describe the mechanism, we divide the system into 
a conductive electron system and a magnetic subsystem of localized spins. 
We introduce an antiferromagnetic exchange coupling between the localized 
spins, and Kondo coupling between the electrons and the spins. 
This model is called a generalized Kondo lattice model and 
a spin fermion model. 
We consider a multilayer structure and parallel magnetic fields, 
in which the orbital pair-breaking effect is suppressed.

When the magnetic field is applied to the antiferromagnetic 
long-range order, a canted spin structure occurs. 
Then, the ferromagnetic moment is induced in the spin subsystem, 
and it creates the exchange field on the mobile electrons, 
which cancels the magnetic field in the Zeeman energy term.

In addition to the exchange fields, 
the localized spins create the internal magnetic fields, 
which cause the Lorentz force in the mobile electrons. 
We ignore this additional internal field for simplicity 
and to concentrate on the spin effect. 
This simplification is quantitatively correct in the case where 
magnetic layers and conductive layers are spatially separated. 
For example, in \RSGCO \hsp{0.25} compounds, the ruthenate layers are distant 
from the cuprate layers, and the internal magnetic field is considered 
to be weak~\cite{Pic99}.

We assume the long-range order for simplicity, 
but the same effect is expected for the short-range order. 
Within the antiferromagnetic correlation length, 
the cancellation effect occurs. 
The fluctuations of the correlated spins do not change the direction of the 
exchange field, which is always opposite to the magnetic field.

First, we consider canted spin structures of the localized spin system. 
In reality, the localized spin state is modified 
by the mobile electron state, 
and the total state should be determined 
self-consistently~\cite{Ham95}. 
However, we assume that such modifications are already included in 
the spin Hamiltonian considered below, 
and that the canted spin structure exists as a renormalized state. 
These assumptions are qualitatively appropriate for the systems 
with strong antiferromagnetic correlations.

The spin Hamiltonian is defined by 
\def\eqSpinH{(1)}
$$
     {\cal H}_J 
     = 
          J \sum_{(i,j)} {\vector S}_i \cdot {\vector S}_j 
          - 2 \mu_0 \sum_i {\vector S}_i \cdot {\vector H} , 
     \eqno\eqSpinH
     $$
where ${\vector H}$ is the magnetic field. 
Here, $\mu_0$ is the magnetic moment of the electron, such that 
$2 \mu_0 = - g \mu_{\rm B} < 0$ 
with the Bohr magneton $\mu_{\rm B}$ and the $g$-factor. 
We define sublattices $A$ and $B$. 
The spin coordinate is defined in two ways for convenience. 
We take the $z$-axis in the direction perpendicular to the layers, 
and the $x$ and $y$-axes parallel to the layers. 
For convenience in consideration of the localized spin states, 
we take $z',x',y'$-axes along the $x,z,y$-axes, respectively. 
We consider the magnetic field applied in the $-x$ direction 
parallel to the layers, that is ${\vector H} = (H,0,0)$ 
with $H < 0$.

We assume the classical spins of the magnitude $S$ for simplicity. 
This treatment is equivalent to the mean field theory at low temperatures, 
when $S$ is replaced by the spin moment. 
We define the polar coordinate $(\theta'_i,\varphi'_i)$ 
to express the direction of the classical spin on site $i$ 
in the spin coordinate system $x'y'z'$, that is, 
$\langle {\vector S}_i \rangle 
  = S \, (\sin \theta'_i \cos \varphi'_i, 
        \sin \theta'_i \sin \varphi'_i, 
        \cos \theta'_i)$. 
When we consider quantum spins, the influence of the shrinkage of 
the spins due to the fluctuations can be partially taken into account 
by the replacement of $S$ with $M = |\langle {\vector S}_i \rangle|$. 
The energy of the localized spin part 
$\langle {\cal H}_J \rangle \equiv E_J$ 
is written as 
\def\eqEJ{(2)}
$$
     \begin{array}{rcl}
     E_J \hsp{-0.5} & = & \hsp{-0.5} \dps{ 
          J S^2 \sum_{(i,j)} (\sin \theta_i \sin \theta_j' 
                               \cos (\varphi_i' - \varphi_j')
               }\\
         && \dps{ 
                               + \cos \theta_i' \cos \theta_j') 
           - 2 \mu_0 H S 
             \sum_{i} \cos \theta_i' , 
               }
     \end{array}
     \eqno\eqEJ
     $$
which becomes minimum 
when $\varphi'_i = \varphi'_j + \pi$ for $i \in A$ and $j \in B$, 
and $\theta'_i = \theta$ for any $i$, 
where $\theta$ is an angle such that 
\def\eqcostheta{(3)}
$$
     \cos \theta \equiv \frac{\mu_0 H}{zJS} 
     \eqno\eqcostheta
     $$
for $|\mu_0 H| \leq zJS$, and $\theta = 0$ for $|\mu_0 H| > zJS$. 
The minimum value is 
$
     E_J = - z N_{\rm s} J S^2 
             [ 1 + 2 (\mu_0 H/zJS)^2 ] / 2 
     $ 
for $|\mu_0 H| \leq zJS$, and 
$
     E_J = - z N_{\rm s} J S^2 
             [ - 1 + 4 \mu_0 H/zJS ] / 2 
     $
for $|\mu_0 H| > zJS$. 
Here, $z$ and $N_{\rm s}$ are the number of nearest neighbor sites 
and the total number of lattice sites, respectively.

Next, we consider the effective Hamiltonian for the mobile electrons, 
which is defined by 
\def\eqelectronH{(4)}
$$
     {\cal H}_e = {\cal H}_0 + {\cal H}_{\rm K}
     \eqno\eqelectronH
     $$
with 
\def\eqHzero{(5)}
$$
     \begin{array}{rcl}
     {\cal H}_0 
     & = & \dps{ 
          - t \sum_{(i,j),\sigma} 
              c_{i\sigma}^{\dagger} c_{j\sigma}
          - \mu \sum_{i,\sigma} 
              c_{i\sigma}^{\dagger} c_{i\sigma} 
              }\\
     && \dps{ 
          - \sum_{i,\sigma_1,\sigma_2} 
              \mu_0 {\vector H} \cdot 
              (c_{i\sigma_1}^{\dagger} 
              {\vec \sigma}_{\sigma_1 \sigma_2} 
              c_{i\sigma_2}) 
           }\\
     \end{array}
     \eqno\eqHzero
     $$
and 
\def\eqHK{(6)}
$$
     \begin{array}{rcl}
     {\cal H}_{\rm K}
     & = & \dps{ 
          J_{\rm K} 
          \sum_{i,\sigma_1,\sigma_2} 
          {\vector S}_i 
          \cdot 
          (c_{i\sigma_1}^{\dagger} 
          {\vec \sigma}_{\sigma_1 \sigma_2} 
          c_{i\sigma_2}) . 
           }\\
     \end{array}
     \eqno\eqHK
     $$
Here, we have omitted pairing interaction terms 
in eq.~{\eqelectronH}~\cite{Shi94c}, 
since our purpose is to examine the modification of 
the electron dispersion relation by the exchange fields. 
In terms of the interlayer hopping energy $t_{\perp}$ and the on-site Coulomb 
repulsion on the magnetic layers $U_{\rm m}$, 
the Kondo coupling is written as $J_{\rm K} = t_{\perp}^2/U_{\rm m} > 0$ 
within a second-order perturbation theory.

In the background of the canted spin structure of the magnetic layers, 
${\cal H}_{\rm K}$ is rewritten as 
\def\eqHKinab{(7)}
$$
     \begin{array}{rcl} 
     \lefteqn{ {\cal H}_{\rm K} = \dps{ 
                 J_{\rm K} S \cos \theta 
                 \Bigl [
                       \sum_{i \in A} (
                            a_{i\uparrow}^{\dagger} a_{i \downarrow} 
                          + a_{i\downarrow}^{\dagger} a_{i \uparrow} ) 
                     }}\\[12pt]
               && \dps{ 
                     + \sum_{j \in B} (
                            b_{j \uparrow}^{\dagger} b_{j \downarrow} 
                          + b_{j \downarrow}^{\dagger} b_{j \uparrow} ) 
                  \Bigr ]
                 + 
                 J_{\rm K} S \sin \theta 
                     }\\[12pt]
               && \dps{ 
                 \times 
                 \Bigl [
                       \sum_{i \in A, \sigma} 
                            \sigma a_{i\sigma}^{\dagger} a_{i \sigma} 
                     - \sum_{j \in B, \sigma} 
                            \sigma b_{j\sigma}^{\dagger} b_{j \sigma} 
                  \Bigr ] , 
                     }\\
     \end{array}
     \eqno\eqHKinab
     $$
where we have defined the electron operators 
$a_{i\sigma}$ and $b_{j\sigma}$ on the sites $i \in A$ and $j \in B$. 
Here, we find that the weak ferromagnetic moment in the $x$ direction 
($S \cos \theta > 0$) increases the population of spins 
in the $-x$ direction when $J_{\rm K} > 0$. 
Furthermore, ${\cal H}_0$ is also rewritten as 
\def\eqHzeroinab{(8)}
$$
     \begin{array}{rcl}
     \lefteqn{ {\cal H}_0 = \dps{ 
          - \, t \dblesum{(i,j),\sigma}{i \in A, j \in B} 
            (a_{i\sigma}^{\dagger} b_{j\sigma} 
            + b_{j\sigma}^{\dagger} a_{i\sigma})
            - \mu \Bigl [
                \sum_{i \in A,\sigma} a_{i\sigma}^{\dagger} a_{i\sigma} 
               }}\\[24pt]
         && \dps{ 
              + \sum_{j \in B,\sigma} b_{j\sigma}^{\dagger} b_{j\sigma} 
                  \Bigr ]
            - \mu_0 H 
                 \Bigl [
                       \sum_{i \in A} (
                            a_{i\uparrow}^{\dagger} a_{i \downarrow} 
                          + a_{i\downarrow}^{\dagger} a_{i \uparrow} ) 
          }\\[16pt]
         && \dps{ 
                     + \sum_{j \in B} (
                            b_{j \uparrow}^{\dagger} b_{j \downarrow} 
                          + b_{j \downarrow}^{\dagger} b_{j \uparrow} ) 
                  \Bigr ] . 
          }\\
     \end{array}
     \eqno\eqHzeroinab
     $$
Therefore, we obtain an expression 
\def\eqHcondeleinab{(9)}
$$
     \begin{array}{rcl}
     {\cal H}_{\rm e} \hsp{-0.5} & = & \hsp{-0.5} \dps{
            {\sum_{\vk}}'
            \left ( 
                    \begin{array}{cccc}
                    \hsp{-0.5} 
                    a_{\vk \uparrow}^{\dagger}   
                    \hsp{-0.5} & \hsp{-0.5} 
                    b_{\vk \uparrow}^{\dagger} 
                    \hsp{-0.5} & \hsp{-0.5} 
                    a_{\vk \downarrow}^{\dagger} 
                    \hsp{-0.5} & \hsp{-0.5} 
                    b_{\vk \downarrow}^{\dagger} 
                    \end{array}
            \right ) 
            {\cal E_{\vk}} 
            \left ( 
                    \begin{array}{c}
                    a_{\vk \uparrow}   \\
                    b_{\vk \uparrow}   \\
                    a_{\vk \downarrow} \\
                    b_{\vk \downarrow} \\
                    \end{array}
            \right ) 
          }
     \end{array}
     \eqno\eqHcondeleinab
     $$
with $4 \times 4$ matrix ${\cal E_{\vk}}$ defined by 
\def\eqcalEk{(10)}
$$
     {\cal E_{\vk}} 
     = \left (
       \begin{array}{cccc}
       - \mu - h_z     & \epsilon_{\vk} & - h_x           & 0               \\
       \epsilon_{\vk}  & - \mu + h_z    & 0               & - h_x           \\
       - h_x           & 0              & - \mu + h_z     & \epsilon_{\vk}  \\
       0               & - h_x          & \epsilon_{\vk}  & - \mu - h_z     \\
       \end{array}
       \right ) 
     \eqno\eqcalEk
     $$
and 
\def\eqdefinitions{(11)}
$$
     \begin{array}{rcl}
     \epsilon_{\vk} & \equiv  & - 2 t \, (\cos k_x + \cos k_y)             \\
     h_x            & \equiv  & \mu_0 H - J_{\rm K} S \cos \theta   \\
     h_z            & \equiv  & - J_{\rm K} S \sin \theta ,                \\
     \end{array}
     \eqno\eqdefinitions
     $$
where $\vk$ and ${\sum_{\vk}}'$ are the sublattice momentum 
and the summation over the first Brillouin zone of 
the sublattice momentum space, respectively. 
The lattice constants are taken as unity.

The eigenvalues of the matrix ${\cal E_{\vk}}$ are expressed as 
\def\eqtildeEk{(12)}
$$
     {\tilde \epsilon}_{\vk} = - \mu \pm 
     \Bigl [
          (\epsilon_{\vk} \pm h_x )^2 + h_z^2 
     \Bigr ]^{1/2} . 
     \eqno\eqtildeEk
     $$
If the system is away from the half-filling and 
has any pairing interactions, superconductivity occurs 
in the renormalized electron system with the energy dispersion relation 
described by eq.~{\eqtildeEk}~\cite{Shi94b}.

The first plus or minus sign $\pm$ in eq.~{\eqtildeEk} corresponds to 
the upper and lower bands which are divided by the exchange field 
in the $z$ direction. 
The band gap does not affect the superconducting transition when the system 
is away from the half-filling, 
whereas near the half-filling, 
the mobile electron layer becomes an insulator 
and the superconductivity does not occur when $\theta \ne 0$, 
i.e., $|\mu_0 H| < zJS$.

The second plus or minus sign $\pm$ corresponds to the split of the Fermi surfaces 
of up and down spin electrons 
when the $x$-axis is taken as the quantization axis of the spin space. 
This split causes the pair-breaking effect, and as a consequence 
the upper critical field is bounded by the Pauli paramagnetic limit. 
However, as shown in eq.~{\eqdefinitions}, 
the magnitude of $h_x$ can be smaller than $|\mu_0 H|$ 
due to the cancellation of $\mu_0 H$ and the exchange field 
$J_{\rm K}S\cos\theta$. 
From eqs.~{\eqcostheta} and {\eqdefinitions}, 
the effective field $H_{\rm eff}$, 
such that $h_x = \mu_0 H_{\rm eff}$, is 
\def\eqHeff{(13)}
$$
     H_{\rm eff} = H \, 
     {\Bigl (} 1 - \frac{J_{\rm K}}{z J} 
     {\Bigr )} 
     \eqno\eqHeff
     $$
for $|\mu_0 H| \leq zJS$, 
and $H_{\rm eff} = H - {\rm sign}(H) J_{\rm K}S/|\mu_0|$ 
for $|\mu_0 H| > zJS$.

The superconductivity occurs when $|H_{\rm eff}| < H_{c}^{(0)}(T)$, 
where $H_{c}^{(0)}(T)$ is the upper critical field ($\HP$ or an FFLO 
critical field) 
in the absence of the exchange field at a temperature $T$. 
Here, we assume that the antiferromagnetic transition occurs at a 
higher temperature, and the magnetic order can be regarded as a 
rigid background. 
Therefore, for $|\mu_0 H| \leq zJS$, the superconductivity occurs when 
\def\eqHc{(14)}
$$
     |H| < H_c 
     \equiv \frac{H_c^{(0)}}{|{1-J_{\rm K}/zJ}|} . 
     \eqno\eqHc
     $$ 
Thus, $H_c$ is the critical field 
when $J_{\rm K} S < zJS - \mu_0 H_c^{(0)}$ 
or $J_{\rm K} S > zJS + \mu_0 H_c^{(0)}$. 
On the other hand, for $|\mu_0 H| \geq zJS$, 
the superconductivity occurs when 
\def\eqHchigh{(15)}
$$
     \frac{J_{\rm K}S}{|\mu_0|} - H_c^{(0)} 
     < |H| < 
     \frac{J_{\rm K}S}{|\mu_0|} + H_c^{(0)} . 
     \eqno\eqHchigh
     $$

\begin{figure}[htb]
\begin{center}
\leavevmode \epsfxsize=7cm  
\epsfbox{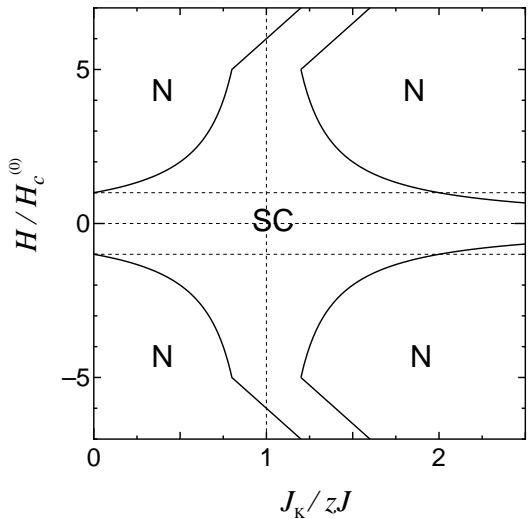}
\end{center}
\caption{
Phase diagram on the $J_{\rm K}/zJ$ and $H/H_c^{(0)}$ 
plane when $|\mu_0|H_{c}^{(0)}/zJS= 0.2$. 
The signs N and SC denote the normal state and the superconductivity, 
respectively. 
} 
\label{fig:phased}
\end{figure}

Therefore, we obtain the phase diagram shown in Fig.~\ref{fig:phased}. 
We have chosen the parameter $|\mu_0|H_c^{(0)}/zJS = 0.2$ 
as an example, which is consistent with our assumption that the magnetic 
long-range or short-range order exists at the superconducting transition 
temperature. 
It is found that when $J_{\rm K}S < zJS - |\mu_0|H_c^{(0)}$ 
the upper critical field is enhanced by the factor $1/(1-J_{\rm K}/zJ)>1$, 
but a field-induced transition is not obtained, 
in contrast to the Jaccarino and Peter mechanism. 
When $J_{\rm K} \sim zJ$, 
the critical field reaches a value of the order of 
$H_c \sim zJS/|\mu_0| \sim T_{\rm AF}^{*}/|\mu_0|$. 
Here, $T_{\rm AF}^{*}$ is a crossover temperature 
at which antiferromagnetic fluctuations begin to occur. 
In the absence of low-dimensional thermal fluctuations, 
$T_{\rm AF}^{*}$ is of the order of 
the antiferromagnetic transition temperature.

For strong Kondo coupling 
$J_{\rm K}S > zJS + |\mu_0|H_c^{(0)}$, 
we find a reentrant transition to a high-field phase 
in addition to a low-field phase. 
For the high fields $|\mu_0H| > zJS$, the present mechanism 
coincides with the original Jaccarino and Peter mechanism, 
since the spin moments are saturated.

\vspace{\baselineskip}

In conclusion, we find that the upper critical field is enhanced 
by the coupled antiferromagnetic layers due to a cancellation 
effect of the external field and the exchange field induced by the canted 
spin structure. 
In particular, when $J_{\rm K} \sim zJ$, 
the system is almost free from the paramagnetic pair-breaking effect 
for practical strengths of the magnetic field. 
The critical field reaches a value of the order of 
$T_{\rm AF}^{*}/|\mu_0|$. 
If such an antiferromagnetic quasi-two-dimensional metal is synthesized 
and exhibits superconductivity, 
it can be a superconductor with an extraordinarily high critical field.

The present phase diagram is very different from that of the orignial 
Jaccarino and Peter mechanism for ferromagnetic metals. 
It is found that 
for small $J_{\rm K}$, a field-induced superconductivity does not occur, 
whereas for large $J_{\rm K}$, 
both the field-induced superconductivity 
and the low-field superconductivity occur.

In the phase diagram, the metal-insulator transition is not presented. 
However, for example, at the half-filling, 
the system is an insulator with a canted spin structure 
when $|\mu_0 H| < zJS$. 
In this case, the phase diagram for large $J_{\rm K}$ 
is similar to that in the $\lambda$-\BETSFeCl \hsp{0.25} 
compound~\cite{Uji01,Note01}. 
The spin moments on ${\rm FeCl_4}$ would create an exchange field on the 
conduction hole band of the two-dimensional network of the ${\rm BETS}$ 
molecules. 
The hybrid ruthenate cuprate compounds are also possible candidates 
due to their crystal structures.

This work was suppoted by a grant for Core Research for Evolutionary 
Science and Technology form Japan Science and Technology Corporation.




\begin{thebibliography}{99}
\bibitem{Shi94b} 
  In this paper, we consider a situation in which the coherent part of 
  the electron excitation remains after the renormalization 
  by the spin wave.
\end{thebibliography}
\end{document}